# ADVANCED CLOUD PRIVACY THREAT MODELING

## Ali Gholami and Erwin Laure

HPCViz Department,
KTH Royal Institute of Technology, Stockholm, Sweden
`{gholami,erwinl@pdc.kth.se}`

## ABSTRACT

*Privacy-preservation for sensitive data has become a challenging issue in cloud computing. Threat modeling as a part of requirements engineering in secure software development provides a structured approach for identifying attacks and proposing countermeasures against the exploitation of vulnerabilities in a system. This paper describes an extension of Cloud Privacy Threat Modeling (CPTM) methodology for privacy threat modeling in relation to processing sensitive data in cloud computing environments. It describes the modeling methodology that involved applying Method Engineering to specify characteristics of a cloud privacy threat modeling methodology, different steps in the proposed methodology and corresponding products. We believe that the extended methodology facilitates the application of a privacy-preserving cloud software development approach from requirements engineering to design.*

## 1. INTRODUCTION

Many organizations that handle sensitive information are considering using cloud computing as it provides easily scalable resources and significant economic benefits in the form of reduced operational costs. However, it can be complicated to correctly identify the relevant privacy requirements for processing sensitive data in cloud computing environments due to the range of privacy legislation and regulations that exist. Some examples of such legislation are the EU Data Protection Directive (DPD) [1] and the US Health Insurance Portability and Accountability Act (HIPAA) [2], both of which demand privacy-preservation for handling personally identifiable information.

Threat modeling is an important part of the process of developing secure software – it provides a structured approach that can be used to identify attacks and to propose countermeasures to prevent vulnerabilities in a system from being exploited [3]. However, the issues of privacy and security are really two distinct topics [4] as security is a core privacy concept, and the current focus of the existing threat modeling methodologies is not on privacy in cloud computing, which makes it difficult to apply these methodologies to developing privacy-preserving software in the context of cloud computing environments.

In 2013, the Cloud Privacy Threat Modeling (CPTM) [6] methodology was proposed as a new threat modeling methodology for cloud computing. The CPTM approach was originally designed to support only the EU DPD, for reducing the complexity of privacy threat modeling. Additionally, there were weaknesses in threat identification step through architectural designs in the early stages of Software Development Life Cycle (SDLC) that demanded improvements.

This paper describes an extension of the CPTM methodology according to the principles of Method Engineering (ME) [5]. The method that has been applied is one known as "Extension-based", which is used for enhancing the process of identifying privacy threats by applying meta-models/patterns and predefined requirements. This new methodology that is being proposed provides strong methodological support for privacy legislation and regulation in cloud computing environments. We describe the high-level requirements for an ideal privacy threat modeling methodology in cloud computing, and construct an extension of CPTM by applying the requirements that were identified.

The rest of this paper is organized as follows. Section 2 provides a background to these developments by outlining the CPTM methodology and existing related work. Section 3 describes the characteristics that are desirable in privacy threat modeling for cloud computing environments. Section 4 describes the steps and products for the proposed new methodology. Section 5 presents the conclusions from this research and directions for future research.

## 2. BACKGROUND AND RELATED WORK

The CPTM [6] methodology was proposed as a specific privacy-preservation threat modeling methodology for cloud computing environments that process sensitive data within the EU's jurisdiction. The key differences between the CPTM methodology and other existing threat modeling methodologies are that CPTM provides a lightweight methodology as it encompasses definitions of the relevant DPD [1] requirements, and in addition that it incorporates classification of important privacy threats, and provides countermeasures for any threats that are identified.

For the first step in the CPTM approach, the DPD terminology is used to identify the main entities to cloud environments that are in the process of being developed. Secondly, the CPTM methodology describes the privacy requirements that must be implemented in the environment, e.g., lawfulness, informed consent, purpose binding, data minimization, data accuracy, transparency, data security, and accountability. Finally, the CPTM approach provides countermeasures for the identified threats. Detailed description of the CPTM methodology steps have been discussed in [6] (Sections 3, 4 and 5).

While the CPTM methodology was the first initiative for privacy threat modeling for cloud computing environments in accordance with the EU's DPD, it nevertheless does not support other privacy legislation, such as that required under the HIPAA [2]. In this paper, we identify the CPTM methodology weaknesses such as support for different privacy legislation and threat identification process and refine the methodology by applying an Extension-based ME approach.

There has been a significant amount of research in the area of threat modeling for various information systems with the goal of identifying a set of generic security threats [7], [8], and [9]. There are guidelines for reducing the security risks associated with cloud services, but none of these include an outline of privacy threat modeling. The Cloud Security Alliance (CSA) guidelines [10] are not thorough enough to be referred as a privacy threat model because they are not specific to privacy-preservation.

The European Network and Information Security Agency (ENISA) has identified a broad range of both security risks and benefits associated with cloud computing, including the protection of sensitive data [11]. Pearson [4] describes the key privacy challenges in cloud computing that arise from a lack of user control, a lack of training and expertise, unauthorized secondary usage, complexity of regulatory compliance, trans-border data flow restrictions, and litigation.

LINDDUN [12] is an approach to privacy modeling that is short for "likability, identifiability, non-repudiation, detectability, information disclosure, content unawareness, and non-compliance". This approach proposes a comprehensive generic methodology for the elicitation of privacy requirement through mapping initial data flow diagrams of application scenarios to the corresponding threats. The Commission on Information Technology and Liberties (CNIL) has proposed a methodology for privacy risk management [13] that may be used by information systems that must comply with the DPD.

# 3. CHARACTERISTICS OF A PRIVACY THREAT MODELING METHODOLOGY FOR CLOUD COMPUTING

This section describes the features that we believe a privacy threat model should have in order to be used for developing privacy-preserving software in clouds in an efficient manner. Based on the properties that are identified, we then apply the Extension-based methodology design approach to construct an extension of the CPTM for supporting various privacy legislation in Section 4.

## 3.1. Privacy Legislation Support

Methodological support for the regulatory frameworks that define privacy requirements for processing personal or sensitive data is a key concern. Privacy legislation and regulations can become complicated for cloud customers and software engineering teams, particularly because of the different terminologies in use in the IT and legal fields. In addition, privacy threat modeling methodologies are not emphasized in existing threat modeling methodologies, which causes ambiguity for privacy threat identification.

## 3.2. Technical Deployment and Service Models

Cloud computing delivers computing software, platforms and infrastructures as services based on pay-as-you-go models. Cloud service models can be deployed for on-demand storage and computing power can be provided in the form of software-as-a-service (SaaS), platform-as-a-service (PaaS) or infrastructure-as-a-service (IaaS) [14]. Cloud services can be delivered to consumers using different cloud deployment models: private cloud, community cloud, public cloud, and hybrid cloud. Table 1, outlines the five essential characteristics of cloud computing [14].

### 3.3. Customer Needs

The actual needs of the cloud consumers must be taken into consideration throughout the whole life cycle of a project. Additionally, during the course of a project, requests for changes often arise and these may affect the design of the final system. Consequently it is important to identify any privacy threats arising from the customer needs that result from such change requests. Customer satisfaction can be achieved through engaging customers from the early stages of threat modeling so that the resulting system satisfies the customer's needs while maintaining adequate levels of privacy.

### 3.4. Usability

Cloud-based tools aim at reducing IT costs and supporting faster release cycles of high quality software. Threat modeling mechanisms for cloud environments should therefore be compatible with the typical fast pace of software development in clouds-based projects. However producing easy-to-use products with an appropriate balance between maintaining the required levels of privacy while satisfying the consumer's demands can be challenging when it comes to cloud environments.

### 3.5. Traceability

Each potential threat that is identified should be documented accurately and be traceable in conjunction with the associated privacy requirements. If threats can be traced in this manner, it means that threat modeling activities are efficient in tracing of the original privacy requirements that are included in the contextual information and changes over the post-requirement steps such as design, implementation, verification and validation.

Table 1, The five essential characteristics of cloud computing [14]

| Cloud Characteristic | Description | Application |
|---|---|---|
| On-demand self-service | For automatically providing a consumer with provisioning capabilities as needed. | Server, Time, Network and Storage |
| Broad network access | For heterogeneous thin or thick client platforms. | Smartphones, tablets, PCs, wide range of locations |
| Resource pooling | The provider's computing resources are pooled to serve multiple consumers using a multi-tenant model. | Physical and virtual resources with dynamic provisioning |
| Rapid elasticity | Capabilities can be elastically provisioned and released, in some cases automatically, to scale rapidly outward and inward with demand. | Adding or removing nodes, servers, resource or instances |

| **Measured service** | Automated control and optimization of a resource through measuring or monitoring services for various reasons, including billing, effective use of resources, or predictive planning. | Storage, processing, billing, , bandwidth, and active user accounts |
|---|---|---|

# 4. METHODOLOGY STEPS AND THEIR PRODUCTS

Motivated by the facts that privacy and security are two distinct topics and that no single methodology could fit all possible software development activities, we apply ME that aims to construct methodologies to satisfy the demands of specific organizations or projects [17]. In [5], ME is defined as "the engineering discipline to design, construct, and adapt methods, techniques and tools for the development of information systems".

There are several approaches to ME [17], [15] such as a fundamentally "ad-hoc" approach where a new method is constructed from scratch, "paradigm-based" approaches where an existing meta-model is instantiated, abstracted or adapted to achieve the target methodology, "Extension-based" approaches that aim to enhance an existing methodology with new concepts and features, and "assembly-based" approaches where a methodology is constructed by assembling method fragments within a repository.

Figure 1 represents different phases in a common SDLC. Initial security requirements are collected and managed in the requirements engineering phase (A). This includes identifying the quality attributes of the project and assessing the risk associated with achieving them. A design is composed of architectural solution, attack surface analysis and the privacy threat model. Potential privacy threats against the software that is being developed are identified and solutions are proposed to mitigate for adversarial attacks (B). The proposed solution from the design phase is implemented through a technical solution and deployment (C). This includes performing static analysis on source code for software comprehension without actually executing programs. The verification process (D) includes extensive testing, dynamic analysis on the executing programs on virtual resources and fuzzing as a black-box testing approach to discover coding errors and security loopholes in the cloud system. Finally, in the Validation phase the end-users participate to assess the actual results versus their expectations, and may put forth further change requests if needed.

Our proposed methodology identifies the privacy requirements in the Requirements Engineering step, as shown in Figure 2. The results from the Requirements Engineering, which include specifications for privacy regulatory compliance, are fed into the Design step, where activities such as specifying the appropriate cloud environment, identifying privacy threats, evaluating risks and mitigating threats are conducted. Then the produced privacy threat model would be used in the implementation step finally it would be verified and validated in the subsequent steps.

Cloud stakeholders and participants such as cloud users, software engineering team and legal experts will engage in the activities shown in Figure 2 to implement the threat model in context of steps A and B in Figure 1. Cloud software architect as a member of the software engineering team initiates a learning session to clarify the methodology steps and their products, privacy requirements (introducing the law title that is needed to be enforced in the cloud environment), and quality attributes such as performance, usability. The legal experts will identify the definitive requirements that ensure the privacy of data in the platform. In the Design step, the cloud software architect presents architecture of the developing cloud environment for various participants. This will result in a unified terminology to be used in the privacy threat model.

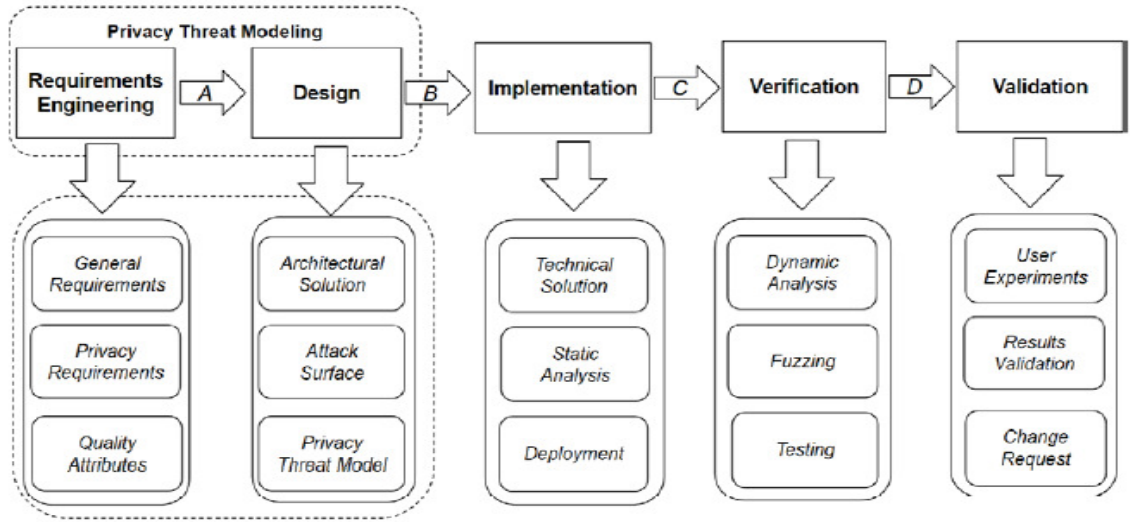

Figure 1, Privacy Threat Modeling in Requirements Engineering and Design of a SDLC

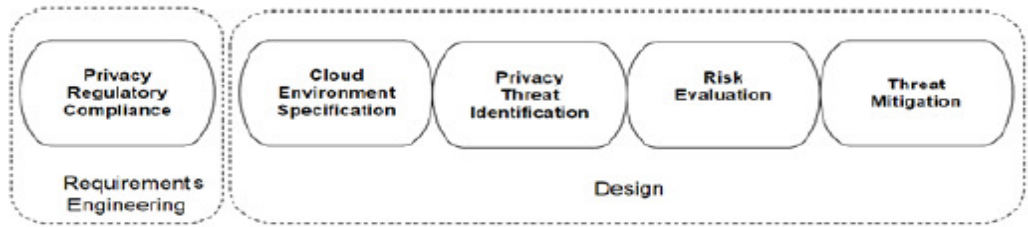

Figure 2, Overview of the Extended CPTM Methodology Steps

The rest of this section outlines the implementation model of the steps represented in Figure 2.

## 3.1 Privacy Regulatory Compliance

Interpreting privacy regulatory frameworks can be often complex for software engineering teams. In the privacy regulatory compliance step, learning sessions with privacy experts, end-users and requirements engineers facilitates the elicitation of privacy requirements (PR). For example, in the EU DPD some of the privacy requirements are: lawfulness, informed consent, purpose binding, transparency, data minimization, data accuracy, data security, and accountability [6]. Each of the requirements that are identified will be labeled with an identifier, e.g., (PRi), name and description to be used in later stages.

## 3.2 Cloud Environment Specification

To ensure that the final cloud software will comply with the relevant legal and regulatory framework, several of the key characteristics that are affected by cloud computing services (including virtualization, outsourcing, offshoring, and autonomic technologies) must be specified.

For this purpose, the physical/logical architectures of the deployment and service model can be developed according to the following steps

- **Step A:** Define the cloud actors [18] (such as ***Cloud Consumer, Cloud Provider, Cloud Auditor, Cloud Broker, and Cloud Carrier***). Cloud consumer is a person or organization uses service from cloud providers in context of a business relationship. Cloud provider makes service available to interested users. Cloud auditor conducts independent assessment of cloud services, operations, performance and security of the deployment. Cloud broker manages the use, performance and delivery of cloud services and establishes relationships between cloud providers and cloud consumers. Cloud carrier provides connectivity and transport of cloud services from cloud providers to cloud consumers through the network.

- **Step B:** Describe a detailed model of the cloud deployment physical architecture where the components will be deployed across the cloud infrastructure. This should give details of where the components will be deployed and run, for example, the operating system version, the database version, the virtual machine location, and where the database server will run.

- **Step C:** Describe the logical architecture of the cloud services model where the major cloud services, along with and the relationships between them that are necessary to fulfill the project requirements, are recorded. This should include the data flow and connections between the relevant cloud services and actors. Note that in this context, an entity is a cloud service with a set of properties that meet a specific functional requirement.

- **Step D:** Describe the assets that need to be protected, the boundaries of the cloud and any potential attackers that might endanger either the cloud environment or the assets that have been identified as being associated with that particular cloud.

The cloud environment specification step consists of composing an architectural report including assets that are subject to privacy protection, cloud actors, physical architecture of the deployment model, and logical architecture of the service model.

### 3.3 Privacy Threat Identification

In this step, privacy threats against the PRs that were established in section 3.1 will be identified and analyzed. To achieve this, the system designers will undertake the following steps.

- **Step A:** Select a privacy requirement from the PR list for threat analysis, e.g., (PR2).

- **Step B:** Correlate identified cloud actors (Step A from Section 3.2) with the actor roles that are defined in the project's privacy law. For example, correlating ***the Data Controller role as a Cloud Consumer, or the Data Processor role as a Cloud Provider*** in the DPD.

- **Step C:** Identify all the technical threats that can be launched by an adversary to privacy and label them in the specified cloud environment. Each identified threat can be named as a T$(i,j)$, where i indicates that threat T that corresponds to PR$i$ and $j$ indicates the actual threat number. For example, in T(2,5) 2 indicates relevance of the threat to PR2 and 5 is the actual threat number.

- **Step D:** Repeat the previous steps until all PRs are processed.

The threat identification step consists of composing an analysis report including a list of threats including id, name, date, author, threat scenario for each class of the PRs.

## 3.4 Risk Evaluation

In this step, all actors participate to rank the threats that have been identified in Section 3.3 with regard to their estimated level of importance and the expected severity of their effect on the overall privacy of the cloud environment. The Importance indicates the likelihood of a particular threat occurring and the level of the Effect indicates the likely severity of the damage if that threat against the cloud environment were carried out.

Assume there are three identified PRs ($PR_1$, $PR_2$, $PR_3$) in addition to related privacy threats T(*1,4*), T(*2,1*) and T(*3,3*) from previous steps for an imaginary cloud system. In this imagined cloud, various participants in the project such as Alice (Cloud Consumer), Bob (Cloud Provider), Dennis (Software Architect), Tom (Lawyer) and Rosa (Cloud Carrier) evaluate the corresponding risk of each identified threats, as illustrated in Table 2.

Table 2, Prioritization of the identified threats, L (Low), M (Moderate), H(High)

| ID | Name | Scenario | Importance | Effect | Participants |
|---|---|---|---|---|---|
| T(*1,4*) | Data Accumulation over Time | The cloud system stores a huge amount of data from Cloud Consumers over the time. This can be done through extensive analysis over collected data from different sources. | H | M | Alice, Bob, Dennis, Tom |
| T(*2,1*) | Linkability of Records | A record owner can be linked through the adversarial background knowledge for the published data to the Cloud Provider. | H | H | Alice, Tom, Bob |
| T(*3,3*) | Cross-linking of data processing | A Cloud Consumer is able to run cross-linking queries over multiple data sets from different data sources. | M | H | Tom, Bob, Rosa |

This step results in composing a risk evaluation report similar to the example in Table 2. This report prioritizes the importance and effects of the privacy threats and it will be used in the Threat Mitigation step in Section 3.5.

## 3.5 Threat Mitigation

In this step, the threat modeling team propose countermeasures to the threats that were identified in the previous step as having the highest likelihood of occurrence and the worst potential effects on the cloud environment. Each countermeasure should clearly describe a solution that reduces the probability of the threat occurring and that also reduces the negative effects on the cloud if the threat was carried out.

Finally, the recommended countermeasures from this step should be documented and fed into the implementation step to be realized through coding and for their effectiveness to be assessed by static analysis. In the later stages of verification and validation, each such countermeasure will be evaluated and approved by the participants.

# 5. CONCLUSIONS AND FUTURE WORK

In this paper we identified the requisite steps to build a privacy threat modeling methodology for cloud computing environments using an Extension-based Method Engineering approach. For this purpose, we extended the Cloud Privacy Threat Modeling (CPTM) methodology to incorporate compliance with various legal and regulatory frameworks, in addition to improving the threat identification process.

In future research, we aim to apply the proposed methodology within domain independent clouds that process sensitive data. This will validate our methodology for providing customized privacy threat modeling for other privacy regulations, such as HIPAA, in cloud computing environments.

## ACKNOWLEDGEMENTS

This work funded by the EU FP7 project Scalable, Secure Storage and Analysis of Biobank Data under Grant Agreement no. 317871.

## AUTHORS

**Ali Gholami** is a PhD student at the KTH Royal Institute of Technology. His research interests include the use of data structures and algorithms to build adaptive data management systems. Another area of his research focuses on the security concerns associated with cloud computing. He is currently exploring strong and usable security factors to enable researchers to process sensitive data in the cloud.

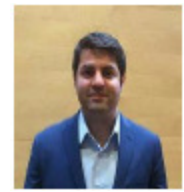

**Professor Erwin Laure** is Director of the PDC - Center for High Performance Computing Center at KTH, Stockholm. He is the Coordinator of the EC-funded "EPiGRAM" and "ExaFLOW" projects as well as of the HPC Centre of Excellence for Bio-molecular Research "BioExcel" and actively involved in major e-infrastructure projects (EGI, PRACE, EUDAT) as well as exascale computing projects. His research interests include programming environments, languages, compilers and runtime systems for parallel and distributed computing, with a focus on exascale computing.

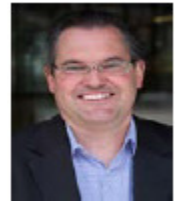